\documentstyle[12pt]{article}
\newcommand{\bea}{\begin{array}{c}}
\newcommand{\eea}{\end{array}}
\newcommand{\beq}{\begin{equation}}
\newcommand{\eeq}{\end{equation}}
\newcommand{\beqa}{\begin{eqnarray}}
\newcommand{\eeqa}{\end{eqnarray}}

\newcommand{\nbar}{\overline{n}}

\newcommand{\rhobar}{\overline{\rho}}
\newcommand{\xibar}{\overline{\xi}}

\newcommand{\gtilde}
	{\mathrel{\raisebox{-1ex}{$\stackrel{\textstyle >}{\sim}$}}}
\newcommand{\ltilde}
	{\mathrel{\raisebox{-1ex}{$\stackrel{\textstyle <}{\sim}$}}}

\newcommand{\lexp}{\mathop{\bigl\langle}}
\newcommand{\rexp}{\mathop{\bigr\rangle}}
\newcommand{\rexpc}{\mathop{\bigr\rangle_c}{}}
\newcommand{\sep}{\mathop{\,;\,}}

\def\etal{{\it et al.\ }}
\def\Or{{\cal O}}

\font\twelveBF=cmmib10 scaled 1200
\newcommand{\x}{\hbox{\twelveBF x}}

\def\km{\,{\rm km}}
\def\s{\,{\rm s}}
\def\Mpc{\,h^{-1}\,{\rm Mpc}}
\newcommand{\ab}{{\{ab\}}}
\newcommand{\eq}{{equation~}}
\newcommand{\Eq}{{Equation~}}

\newcommand{\IO}{{\rm IO}}
\newcommand{\rI}{{\rm I}}
\newcommand{\rO}{{\rm O}}
\font\trm=cmr12

\begin{document}
\baselineskip 24pt

\begin{titlepage}


\vskip 1cm
\begin{center}
{\Large\bf Biasing and Hierarchical Statistics \\
in  Large-scale Structure}
\end{center}

\begin{center}
{\large\sc J. N. Fry\footnote%
{\trm NASA/Fermilab Astrophysics Center,

 Fermi National Accelerator Laboratory, Batavia, IL 60510-0500, USA}
${}^,$
{\trm Department of Physics, University of Florida,

Gainesville, FL 32611, USA}

and Enrique Gazta\~naga${}^1$}

\vskip .4cm
{\small\rm ABSTRACT}
\end{center}

{
In the current paradigm there is a non-trivial bias expected in the

process of galaxy formation.
Thus, the observed statistical properties of the galaxy distribution

do not necessarily extend to the underlying matter distribution.
Gravitational evolution of initially Gaussian seed fluctuations
predicts

that the connected moments of the matter fluctuations exhibit a

hierarchical structure, at least in the limit of small dispersion.
This same hierarchical structure has been found in the galaxy
distribution,

but it is not clear to what extent it reflects properties of the
matter

distribution or properties of a galaxy formation bias.

In this paper we consider the consequences of an arbitrary,
effectively

local biasing transformation of a hierarchical underlying matter
distribution.
We show that a general form of such a transformation preserves the

hierarchical properties and the shape of the dispersion in the limit
of

small fluctuations, i.e. on large scales, although the values of the

hierarchical amplitudes may change arbitrarily.
We present expressions for the induced hierarchical amplitudes

$ S_{g,j} $ of the galaxy distribution in terms of the matter

amplitudes $ S_j $ and biasing parameters for $ j = 3 $--7.
For higher order correlations, $ j > 2 $, restricting to a linear

bias is not a consistent approximation even at very large scales.
To draw any conclusions from the galaxy distribution about matter

correlations of order $j$, properties of biasing must be specified

completely to order $ j-1 $.

}

\bigskip

\noindent
{\it Subject Headings}: Large-scale structure of the universe ---

galaxies: clustering

\end{titlepage}

\baselineskip 15pt

\section{Introduction}

There is accumulating observational evidence that the large scale

galaxy $j$-point correlation functions exhibit a hierarchical
structure.
Averaged over a sphere of radius $R$, this means the order $j$

connected moments obey

\beq
\xibar_j(R)= S_j \, \xibar_2(R)^{j-1},  \label{eq:hibar}
\eeq
where the $S_j$ are constant, independent of $R$.
The hierarchical relation can follow from a scaling symmetry,

$\xi_j(\lambda \x_1, \dots, \lambda \x_j) = \lambda^{-(j-1)\gamma}
\xi_j(\x_1,\dots,\x_j)$, or from the multi-point expression
\beq
\xi_j(\x_1, \dots, \x_j) = \sum_\alpha Q_{j,\alpha}
 \sum_\ab \prod^{j-1} \xi_2(r_{ab})  \label{eq:hier}
\eeq
(Fry 1984$b$).
In the standard graphical notation of field theory, associated with

each term in \eq(\ref{eq:hier}) there is a graph, such that vertices,

or nodes, correspond to the points $ \x_1, \dots, \x_j $, and edges,

or lines, between node $a$ and node $b$ correspond to factors

$ \xi_2(x_{ab}) = \lexp \delta(\x_a) \delta(\x_b) \rexp $

that connect all points.
Thus the hierarchy in \eq(\ref{eq:hier}) is composed of ``tree''
graphs

(connected with no cycles) of $j$ vertices and $j-1$ edges.

The sum over $ \alpha $ denotes topologically distinct graphs; the

sum over $ \ab $ is over relabelings within $ \alpha $.
If all $ Q_{j,\alpha} $ are identical, there are in total $j^{j-2}$
terms,

corresponding to all possible reassignments of the labels $a$,

$ b = 1 $, \dots, $j$, and, up to geometrical factors usually very
close

to 1, $ S_j = j^{j-2} Q_j $.
The hierarchial pattern together with a power-law variance,

$\xibar_2(R) \propto R^{-\gamma}$, is equivalent to a fractal

galaxy distribution.
If $ \xibar_2(R) $ exhibits a change of behavior, as in two power-law
models

(Dekel \& Aarseth 1984; Guzzo \etal 1991; Calzetti, Giavalisco, \&

Meiksin 1992), we might expect the $S_j$ to be constant over the
range

of $R$ where $ \xibar_2 $ has a constant slope, but in practice the

observed $ S_j $ are strikingly constant.
Observations indicate that the hierarchy, \eq(\ref{eq:hibar})

or \eq(\ref{eq:hier}), holds both on mildly linear, $ \xibar_2(R)
\ltilde 1 $,

and nonlinear, $ \xibar_2(R) \gtilde 1 $, scales, at least for the

lower values of $j$, and has been found in angular catalogs of
optical

(e.g. Groth \& Peebles 1977 Fry \& Peebles 1978; Szapudi \etal 1992)
and

IRAS (Meiksin \etal 1992) galaxies.
Similar results have been reported for redshift samples of

IRAS galaxies (Bouchet \etal 1992) and in the CfA and SSRS optical

catalogs (Gazta\~naga 1992; Gazta\~naga \& Yokohama 1993).

Remarkably, this same hierarchical structure is predicted for the
matter
distribution evolved gravitationally in perturbation theory when the

initial fluctuations are Gaussian (e.g. Peebles 1980, Fry 1984$b$,
Goroff

\etal 1986, Bernardeau 1992) and also in the highly nonlinear regime
of

gravitational clustering (Davis \& Peebles 1977, Peebles 1980,

Fry 1984$a$, Hamilton 1988).
But, in order to relate theory with the observations, we have to

address the problem of how well galaxies trace the matter
fluctuations.
Are the observed hierarchical properties of the galaxy distribution a

consequence of the hierarchical properties of matter?

Or, are they an accident or conspiracy of galaxy-matter biasing?
If the galaxy distribution is determined physically by the mass

distribution, then we expect that the number density of galaxies
should be

given as a functional of the mass density, $ n_g(\x) = F[\rho(\x)]$.
Linear biasing, that the galaxy fluctuations are proportional
to the matter fluctuations, $ \delta_g = b \delta_\rho $, is often

assumed as an approximation at large scales.
For this case, up to scalings, all statistical properties are
preserved

by the biasing, and the observed galaxy properties do reflect the
matter

distribution.
However, in the general case, we expect it is highly unlikely that
the

relation is both local and linear.
Below, we study how an arbitrary nonlinear biasing affects
statistical

studies on large scales, $ R \gtilde 10 \Mpc $ (Hubble's constant
$H_0 =

100 \, h \km \s^{-1}\,{\rm Mpc}^{-1} $), where $ \xibar_2(R) \ltilde
1$.
We compute the resulting correlation amplitudes directly for low
order

correlations, and we show that the results extend to all orders.
We argue finally that there may be observational evidence that
biasing

must be a nonlinear transformation, and it is not clear whether
 the linear approximation is good, or even consistent, at large
scales.

\section{Biasing and hierarchical distributions at large scales}

\subsection{One-point statistics}

Let us first consider the statistics for one random variable,

the (smoothed) density contrast $ \delta_W (\x) $:

\beq
\delta_W (\x) = \int d^3x' \, \delta(\x') \, W(\x-\x') ,

\label{eq:window}
\eeq
with $ \delta(\x) = [\rho(\x)-\rhobar]/\rhobar $,

where $ \rho(\x) $ is the local density, $ \rhobar $ the mean density

and $ W(\x) $ a normalized window function.

For a top-hat window, $ \delta_W(\x) $ is just the

volume average of $ \delta(\x) $ over a sphere of radius $R$.
To simplify notation, we use $\delta$ for $\delta_W(\x)$.
The statistical average is over different realizations of $
\delta(\x) $

and corresponds to the average over position in a fair sample of the

universe.

As the result of biasing, we assume that the (smoothed) galaxy
density

can be written as a function of the mass density,

$ \delta_g = [n(\x) - \nbar] / \nbar = f \bigl( \delta \bigr) $,

and express $f$ as a Taylor series:
\beq
\delta_g = f(\delta) = \sum_{k=0}^\infty {b_k \over {k!}} \delta^k.

\label{eq:taylor}
\eeq
The linear term $ b_1 $ corresponds to the usual linear bias factor $
b $.
To have $\lexp \delta_g \rexp = 0 $ we must fix

$ b_0=-\sum_{k=2}^\infty b_k \lexp \delta^k \rexp / k! $.
The value of $ b_0 $ is irrelevant for the connected moments for

$ j \geq 1 $, and we will make no further mention of $ b_0 $.
\Eq(\ref{eq:taylor}) is not the most general possibility; we could

conceive of a relation involving $ \delta $ at all points such as

\beq
\delta_g(\x) = b_0 + \int d^3x' \, b_1(\x') \, \delta(\x - \x')

+ \int d^3x' d^3x''

\, b_2(\x',\x'') \, \delta(\x-\x') \delta(\x-\x'') + \cdots .
\label{eq:nonlocal}
\eeq
However, to lowest order in $ \delta $, \eq(\ref{eq:nonlocal})

would give the two-point function

\beq
\xi'(x_{12}) = \int d^3x'_1 d^3x'_2 \, b(\x'_1) b(\x'_2)

\xi(|\x_{12} - \x'_{12}|) .
\eeq
The observational suggestion that for groups and clusters the
correlations

of selected objects are proportional to those of galaxies,

$ \xi_{cc}(r) = b^2 \xi_{gg}(r) $, is an indication that if the
relation

is nonlocal, the range is relatively short.
For the windowed field (\ref{eq:window}) smoothed over large scales,

\eq(\ref{eq:taylor}) should provide an adequate first approximation.

If the matter density $ \delta $ has hierarchical irreducible
correlations

or cumulants as in \eq(\ref{eq:hier}), we show next that in the limit
of

small $ \xi_2 $, the local biasing transformation in
\eq(\ref{eq:taylor})
 preserves the hierarchical structure.

\subsection{Expressions for the first orders}

For the first few low order correlations we can compute directly

correlations of the biased field in terms of those of the original

matter field and the biasing parameters.
We consider the case of one-point statistics to simplify notation.
We assume that the matter distribution has hierarchical connected
moments

\beq
\xibar_j = \lexp\delta^j\rexpc = S_j~  \xibar_2^{\,j-1} ,

\label{hierarchy1}
\eeq
where $ \xibar_2 = \lexp\delta^2\rexpc $.
We use the generating function method for calculating

$ \lexp \delta_g^{\,j} \rexpc $ from

\beq
\xibar_{g,j} = \lexp\delta_g^{\,j} \rexpc = {d^j \over{dt^j}}

\ln \lexp e^{t \delta_g} \rexp|_{{t=0}} ,

\eeq
(Fry 1985), where the biased field $\delta_g$ is given by

\eq(\ref{eq:taylor}).
This procedure gives the following for $ \xibar_{g,j} $

for $ j = 2 $--5:

\beqa
\xibar_{g,2} &=& b^2 \xibar_2 + b^2 \xibar_2^{\,2}

	(c_2 S_3 + c_3 + c_2^2/2) + \Or(\xibar_2^3) \nonumber \\
\noalign{\smallskip}
\xibar_{g,3} &=& b^3 \xibar_2^{\,2} (S_3 + 3 c_2) + b^3
\xibar_2^{\,3}

	(3 c_2 S_4/2 + 9 c_3 S_3/2 + 6 c_2^2 S_3 + 3 c_4/2 +

	6 c_2 c_3 + c_2^3) + \Or(\xibar_2^{\,4}) \nonumber \\
\noalign{\smallskip}
\xibar_{g,4} &=& b^4 \xibar_2^{\,3}

	(S_4 + 12 c_2 S_3 + 4 c_3 + 12 c_2^2)

	+ b^4 \xibar_2^{\,4} [2 c_2 S_5 + 8 c_3 S_4 + 18 c_2^2 S_4

	+ (6 c_3 + 12 c_2^2) S_3^2 \nonumber \\
&& \qquad

	+ (12 c_4 + 78 c_2 c_3 + 36 c_2^3) S_3

	+ 2 c_5 + 18 c_2 c_4 + 12 c_3^2 + 36 c_2^2 c_3 + 3 c_2^4]
	+ \Or(\xibar_2^{\,5}) , \nonumber \\
\noalign{\smallskip}
\xibar_{g,5} &=& b^5 \xibar_2^{\,4} [S_5 + 20 c_2 S_4 + 15 c_2 S_3^2
+

	(30c_3 + 120 c_2^2) S_3 + 5 c_4 + 60 c_3 c_2+ 60 c_2^3]
\nonumber \\

&& 	+ b^5 \xibar_2^{\,5} [5 c_2 S_6/2 + (25 c_3/2 + 40 c_2^2) S_5

	+ (25 c_3 + 70 c_2^2) S_3 S_4 	+ (25 c_4 + 230 c_2 c_3

	+ 180 c_2^2) S_4

\nonumber \\
&& \qquad

	+ (75 c_4/2 + 330 c_2 c_3 + 240 c_2^3) S_3^2


	+ (25 c_5 + 310 c_2 c_4 + 210 c_2^3 + 1020 c_2^2 c_3 + 240
c_2^4)

	S_3 \nonumber \\
&& \qquad

	+ 5 c_6/2 + 40 c_2 c_5 + 70 c_3 c_4 + 180 c_2^2 c_4 + 240 c_2
c_3^2

	+ 240 c_2^3 c_3 + 12 c_2^5]
	+ \Or(\xibar_2^{\,6}) ,

\label{eq:xibar_g}

\eeqa
where we write $ c_k = b_k / b $ for $ k \geq 2 $.
We have obtained, but do not display, results up to order

$ \Or(\xibar_2^6) $ for $ \xibar_{g,j} $ up to $ j = 7 $.
The leading term in \eq(\ref{eq:xibar_g}) for $ \xibar_{g,2} $ is the

linear bias result, $ \xibar_{g,2} = b^2 \xibar_2 $.
To leading order in $ \xibar_2 $, the remaining results,

$ \xibar_{g,j} $ for $ j \ge 3 $, are hierarchical,

$ \xibar_{g,j}  = S_{g,j} \, \xibar_{g,2}^{\,j-1} $,

with amplitudes $ S_{g,j} $ given by

\beqa
S_{g,3} &=& b^{-1}(S_3 + 3 c_2) \nonumber \\
\noalign{\smallskip}
S_{g,4} &=& b^{-2} (S_4 + 12 c_2 S_3 + 4 c_3 + 12 c_2^2) \nonumber \\
\noalign{\smallskip}
S_{g,5} &=& b^{-3} [S_5 + 20 c_2 S_4 + 15 c_2 S_3^2 +

	(30c_3 + 120 c_2^2) S_3

+ 5 c_4 + 60 c_3 c_2+ 60 c_2^3] \nonumber \\

\noalign{\vfill\eject}
S_{g,6} &=& b^{-4} [S_6 + 30 c_2 S_5 +60 c_2 S_3 S_4 +

  (60 c_3 + 300 c_2^2)S_4  + (90 c_3 + 450 c_2^2) S_3^2 \nonumber \\
&& \qquad + (60 c_4 + 900 c_2 c_3 + 1200 c_2^3) S_3 + 6 c_5

  + 120 c_4 c_2 + 90 c_3^2 + 720 c_3 c_2^2 + 360 c_2^4 ] \nonumber \\
\noalign{\smallskip}
S_{g,7} &=& b^{-5}[S_7 + 42 c_2 S_6 + 105 c_2 S_3 S_5 +

	(105 c_3 + 630 c_2^2) S_5 + 70 c_2 S_4^2 \nonumber \\
&& \qquad + (420 c_3 + 2520 c_2^2) S_3 S_4 +

	(140 c_4 + 2520 c_2 c_3 + 4200 c_2^3) S_4 \nonumber \\
&& \qquad + (105 c_3 + 630 c_2^2 + 315 c_4 + 5670 c_2 c_3

	+ 9450 c_2^3) S_3^2 \nonumber \\
&& \qquad + (105 c_5 + 2520 c_2 c_4 + 1890 c_3^2

	+ 18900 c_2^2 c_3 + 12600 c_2^4) S_3 \nonumber \\
&& \qquad + 7 c_6 + 210 c_2 c_5 + 420 c_3 c_4 + 2100 c_2^2 c_4

	+ 3150 c_2 c_3^2 + 8400 c_2^3 c_3 + 2520 c_2^5]

\label{eq:S_g}

\eeqa
The numerical factors are determined by combinatorics and, as in

perturbation theory, can be related to a counting of tree graphs.
This is especially evident in the terms induced solely by the $ c_k $

(cf. Fry 1984$b$), where the sum of coefficients is just $ j^{j-2} $,

the total number of labeled tree graphs.
Equivalent results were first derived using a different technique by

James \& Mayne (1962), who present contributions up to $ S_6 $, or

$ \Or(\xibar_2^{\,5}) $.
Notice that the parameters $b_j$ in the biasing function can be
chosen

arbitrarily at each order, and thus can modify the matter amplitudes

$S_j$ into arbitrary galaxy amplitudes $S_{g,j}$.

The popular model of bias as a sharp threshold clipping

(Kaiser 1984, Politzer \& Wise 1985, Bardeen \etal 1986, Szalay
1988),

where $ \delta_g = 1 $ for $ \delta > \nu \sigma $ and $ \delta_g = 0
$

otherwise, does not have a series representation

around $ \delta = 0 $.
Such a clipping applied to a Gaussian background produces a
hierarchical

result with $ S_{g,j} = j^{j-2} $ in the limit $ \nu \gg 1 $,

$ \sigma \ll 1 $.
This is the same result as we obtain from \eq(\ref{eq:S_g})

for an exponential biasing of a Gaussian matter distribution,

$ \delta_g = \exp(\alpha \delta/\sigma) $, which is equivalent to the

sharp threshold when the threshold is large and fluctuations are weak

(cf. Bardeen \etal 1986; Szalay 1988).
The exponential bias function has an expansion

$ \delta_g = \sum_k (\alpha \delta / \sigma)^k / k! $

and thus $ c_j = b^{j-1} $, independent of $ \alpha $ and $ \sigma $.
With $ S_j = 0 $, the terms induced by $ c_j = b^{j-1} $ in
\eq(\ref{eq:S_g})

also give $ S_{g,j} = j^{j-2} $.

In a similar way one could compute the multipoint correlations and

the biased multi-point amplitudes $Q_{g,j}$ in terms of the local

matter amplitudes $Q_j$ in \eq(\ref{eq:hier}).
The calculation in this case will be identical to the one for the
smoothed
 fluctuations above, with $S_j$ effectively replaced by $ j^{j-2}Q_j
$,

but with additional attention required for topologically

distinct configurations.

\subsection{General results: One point statistics}

The results summarized in \eq(\ref{eq:S_g}) involve the cancellation

of an increasing number of lower order terms; the raw moments

$ \lexp \delta_g^{\,j} \rexp $ are of order $ \xibar_2^{\,j/2} $.
Thus, that the cumulants of the biased distribution are

also hierarchical is likely to be more than an accident.
This was proved in general in the following theorem by James

(1955) and James and Mayne (1962):
\begin{quote}
{\sc Theorem 1:} If a variate $\delta$ possesses finite cumulants of

all orders with $ \lexp\delta^j\rexpc = \Or (\nu^{-j+1}) $,


and if the cumulants of $ \delta_g = f(\delta) $ are calculated on

the basis of a (possibly formal) Taylor expansion (\ref{eq:taylor})

where the $b_k$ do not depend upon $\nu$, i.e. they are $\Or(\nu^0)$,

then $\lexp\delta_g^j\rexpc = \Or (\nu^{-j+1}) $.
\end{quote}
As noted above, this is by no means obvious for $j > 2$, as the raw

$j$-moment of $\delta_g$ is of order $\nu^{-j/2}$, but on taking the

connected part the terms up to $\Or (\nu^{-j+1})$ always seem to
cancel.
The explanation for this cancelation, i.e. the proof of the theorem,
is based on an adoption of the Fisher rules for obtaining

the sampling cumulants of $k$-statistics (Kendall, Stuart and Ord
1987)
to statistics of polynomial symmetric functions.

To prove the theorem, James (1955) first considers the variables

$ z_r = b_r \delta^r $.
It is straightforward to see that $ \lexp\delta_g^j\rexpc =

\sum_{r_1} \dots \sum_{r_j} \lexp z_{r_1} \cdots z_{r_j} \rexpc $.
Therefore it is sufficient to show that

$ \lexp z_{r_1} \cdots z_{r_j} \rexpc $ is of order $ \nu^{-j+1} $.
Now consider a sample $\delta_1,\dots,\delta_n$
of $n$ independent values of $\delta$ to define the general
statistics
$z_r = b_r (\sum \delta_i)^r $, $ r = 1$, 2, \dots.
James now uses the Fisher rule that states that to find the cumulants

of the $z$-statistics in terms of population cumulants, we can
neglect

an array which splits up into two or more disjoint blocks.
Finally, to conclude the proof, it is necessary to use that each

$\lexp\delta^j\rexp = \Or (\nu^{-j+1}) $; a different structural
relation
is not preserved under the general transformation in
\eq(\ref{eq:taylor}).

This theorem applies directly to the large scale distribution.
{}From the results of perturbation theory we can assume that the
matter distribution, $\delta $, follows the hierarchical relation
$ \xibar_j = \lexp \delta^j \rexpc = S_j \xibar_2^{\,j-1} $ and so we
have the required conditions for the theorem with $ \nu =
\xibar_2^{\,-1}$.
If biasing can be described by a local transformation, so that the
galaxy field $\delta_g$ can be expressed as in \eq(\ref{eq:taylor})
with
$\delta_g = f(\delta)$ then we conclude from the theorem above that

the galaxy distribution will also be hierarchical for small values

of $\xi_2$, i.e. large scales.
Reversely, if the galaxy distribution is hierarchical and if

$ \delta_g = f(\delta)$ then the underlying matter statistics must

be hierarchical at large scales.

\subsection{General results: Multi-point statistics}

James (1955) also considers a more general result using multivariate

sampling rules.
He proposed and proved the following theorem:

\begin{quote}
{\sc Theorem 2.} If $ \delta_{g,1} = f_1(\delta_1, \dots ,\delta_N)
$,

$ \delta_{g,2} = f_2(\delta_1,\dots,\delta_N) $, \dots are functions
of the
variates $ \delta_1 $, \dots, $ \delta_N $ formally expansible in the
forms:
\beq
f_k(\delta_1,\dots,\delta_N) = (b_k)_0 + \sum_i (b_k)_{1,i} \,
\delta_i +
{1\over2} \sum_{i,j} (b_k)_{2,ij} \, \delta_i \delta_j + \cdots ,

\label{taylorm}
\eeq
and if the $j$-cumulant,

$ \lexp \delta_{i_1} \cdots \delta_{i_j} \rexpc = \Or (\nu^{-j+1}) $,
with $i_1,\dots, i_j = 1,\dots, N$ and $j = 1,2,\dots$, then the same

holds for the cumulants $ \lexp \delta_{g,i_1} \cdots
\delta_{g,i_j}\rexpc $

of $\delta_{g,N} $.
\end{quote}

For the case of spatial distribution we can interpret these variates
as
 corresponding to the density contrast at different points,

$ \delta_k = \delta (\x_k) $ and $ \delta_{g,p} = \delta_g(\x_p)$,

so that the multivariate cumulants above are the standard
correlation functions,  $ \lexp \delta(\x_1) \cdots
\delta(\x_j)\rexpc$.
{}From the hierarchy (\ref{eq:hier}) above, the $j$-correlation for
matter

is of order $\nu^{-j+1}$ with $\nu$ the inverse amplitude of the

two-point function.
Therefore a local biasing transformation, \eq(\ref{eq:taylor}):
$ \delta_g(\x_k) = f( \delta(\x_k) )$, will produce

$ \lexp \delta_g (\x_1) \dots \delta_g (\x_j) \rexpc = \Or
(\nu^{-j+1}) $

and consequently the hierarchy (\ref{eq:hier}) for galaxies.

Theorem 2 applies even when the coefficients $ b_k $ are functions

of position, an inhomogeneous, nonlocal biasing transformation,

$\delta_g(x_j) = F[x_j,\delta(x_1),\dots,\delta(x_N)]$.
In this case, the induced correlations can have little in common with
the

underlying matter correlations.
Nevertheless we still have

$\lexp \delta_g(\x_1) \dots \delta_g(\x_j)\rexpc = \Or (\nu^{-j+1})$,

but now with local or scale-dependent values of $Q_{j}$.

\subsection{A bias transformation Group}

In a practical situation, only the lower moments of the

observed galaxy distribution can be determined.
We will define two spatial distributions to be equivalent to order
$N$

if their moments agree up to order $N$; a class of equivalent

distributions will be called an $N$-order distribution.
We can also define the equivalance relation for bias transformations:

two biasing transformations over an $N$-order distribution are
equivalent if, and only if, the first $N$ coefficients of expansion

(\ref{eq:taylor}) are equal.
The set of equivalent classes of transformations will be called

$N$-order biasing or $N$-order transformations.
With this nomenclature, equation (\ref{eq:S_g}) shows
that an $N$-order transformation can arbitrarily change one $N$-order

hierarchical distribution to another.

It is easy to see that $N$-order transformations,
$\{b\sep \dots ~\sep ~c_N\}$, form a {\it non-Abelian Group} of
transformations.

The composition (or group operation) of the transformation

$\{b_B\sep c_{B,2}\sep c_{B,3} \sep \dots ~\}$ following

$\{b_A\sep c_{A,2}\sep c_{A,3} \sep \dots ~\}$ yields the
transformation:
\beq
\{ b_A b_B \sep c_{A,2} +  b_A~ c_{B,2} \sep

c_{A,3} + 3~ b_A \, c_{A,2} \, c_{B,2} + b_A^2 \, c_{B,3} \sep \dots
{}~ \}.

\label{eq:composition}
\eeq
The neutral element is $\{1 \sep 0 \sep \dots \sep 0\}$ and
the inverse is:
\beq
\{b\sep c_2\sep c_3 \sep \dots ~\}^{-1} =
\{ b^{-1} \sep - b^{-1}c_2 \sep b^{-2} (3c_2^2 - c_3) \sep \dots ~\},
\eeq
so that (\ref{eq:S_g}) can be easily inverted to give $S_j$ in terms
of $S_{g,j}$.
These properties will be useful when comparing models with
observations.
For example,  consider that the distribution of both optical (O) and

IRAS (I) selected galaxies are related to the matter distribution by

$ \{ b_\rO \sep c_{\rO,2} \sep c_{\rO,3} \sep \dots ~ \}$ and

$ \{ b_\rI \sep c_{\rI,2} \sep c_{\rI,3} \sep \dots ~ \}$.
Under the group properties, there will also be a biasing
transformation

between the optical and IRAS distributions,

$ \delta_\rI = f_\IO (\delta_\rO) $, with

$ \{ b_{\IO} \sep c_{\IO,2} \sep c_{\IO,3} \sep \dots ~ \} $ given by

\beqa
b_{\IO}   &=& b_{\rI~} /  b_\rO \nonumber \\
\noalign{\smallskip}
c_{\IO,2} &=& b_\rO^{-1} (c_{\rI,2} - c_{\rO,2}) \nonumber \\
\noalign{\smallskip}
c_{\IO,3} &=& b_\rO^{-2}

(c_{\rI,3} - c_{\rO,3}) + 3~b_\rO^{-1}~ c_{\rO,2}~ c_{\IO,2},
\label{eq:IOgroup}
\eeqa
and so on.
Let us apply these properties  to the observations.

\subsection{Biasing between optical and IRAS distributions}

The relations obtained above can be used to fit a phenomenological
bias
between optical (O) and IRAS (I) selected galaxies because,
as pointed out in the introduction, both optical and IRAS
distributions

are hierarchical at large scales, at least to the lower orders.
Direct comparison of the dispersion at different scales gives

$ \lexp \delta^2 \rexp_\rI = b^2_{\IO} \lexp \delta^2\rexp_\rO $,

with $ b_{\IO} = 0.7 \pm 0.1 $ (e.g. Strauss \etal 1992,

Saunders \etal 1992; although this is the value of $b$ quoted at

$R \simeq 8 \Mpc $, there is no significant different for larger
scales),

in agreement with a local biasing transformation.
We will look for a class of transformations

$ \{ b_{\IO} \sep c_{\IO,2} \sep c_{\IO,3} \sep \dots ~ \} $
to relate optical, $\delta_\rO$, and IRAS, $\delta_\rI =
f(\delta_\rO)$
 distributions. From (\ref{eq:S_g}) we have:
\beqa
S_{\rI,3}  &=& b_{\IO}^{-1} (S_{\rO,3} +  3 \, c_{\IO,2}) \nonumber
\\
\noalign{\smallskip}
S_{\rI,4}  &=& b_{\IO}^{-2} (S_{\rO,4} + 12 \, c_{\IO,2} S_{\rO,3}

+ 4\,c_{\IO,3} + 12\,c_{\IO,2}^2).

\label{eq:SIO}

\eeqa
We apply this expression to values for amplitudes found from optical
and

IRAS samples:

\begin{itemize}

\item for optical galaxies: Szapudi \etal (1992) from the Lick sample

obtain  $ S_{\rO,3} = 4.32 \pm 0.21 $ and $ S_{\rO,4} = 31 \pm 5 $.

\item for IRAS galaxies: Meiksin \etal (1992) obtain

$ S_{\rI,3} = 2.19 \pm 0.18 $ and $ S_{\rI,4} = 10.1 \pm 2.9 $.

\end{itemize}
These values are extracted from angular distributions and corrected
for

projection using the same techniques for IRAS and optical galaxies.
We have taken $ S_3 = 3 Q_3 $ and $ S_4 = 16 Q_4 $, and we have

used the results found for the small scale value $ \gamma = 1.8 $.
The hierarchical pattern is an empirical result that does not depend

on a power-law correlation function, but an uncertainty in the value

of $ \gamma $ can affect the inferred amplitudes.
For $ S_3 $, for $ \gamma $ varying from $1.6$ to $1.8$ to $2.0$,

Meiksin \etal find in IRAS that $ S_3 $ changes from

$ S_3 = 2.37 \pm 0.21 $ to $ 2.19 \pm 0.18 $ to $ 1.95 \pm 0.18 $.
However, from their analysis they find $ \gamma = 1.79 \pm 0.07 $.
Thus, it requires a $ 3\sigma $ change in $ \gamma $ to induce a

significant change in the hierarchical amplitudes, and this is a
small

effect compared to other uncertainties.

With these amplitudes and the value of $b_{\IO}$  above we use

\eq(\ref{eq:SIO}) to find obtain

\beqa
c_{\IO,2} &=& -0.93 \pm 0.06 , \nonumber \\
\noalign{\smallskip}
c_{\IO,3} &=& \phantom{-}2.95 \pm 0.65 ,

\eeqa
incompatible, within the estimated errors (added in quadrature)

with a linear biasing between optical and IRAS distributions, which

would imply $ c_{\IO,2} = c_{\IO,3} = 0 $.
By using the group composition properties (\ref{eq:IOgroup}) one can

conclude that $ c_{\rO,2} \neq c_{\rI,2} $, so that both can not be

zero at the same time.
Thus, a linear biasing from matter for both optical and IRAS galaxies
is

inconsistent with the observations cited.

\section{Discussion}

We do not observe the full matter distribution, but at best just part

of the visible galaxy distribution, and, as shown above, in designing
a

bias prescription one must address the problem beyond linear order

to extract meaningful information from higher order galaxy
correlations.
We can think of several distinct stages where a nonlinear processing

may enter between one and the next.
First, the matter field evolves gravitationally from initial
conditions,

a process that is well known to be nonlinear and that from Gaussian

seed fluctuations produces hierarchical statistics, as in

\eq(\ref{eq:hibar}) or \eq(\ref{eq:hier})

(cf. Fry 1984$b$, Goroff \etal 1986, Bernardeau 1992).
This alone guarantees that the matter distribution is not Gaussian.
At some point, physics determines how matter is processed into
candidates

for observation, luminous stars, galaxies, and so on.
Evidence for dark matter suggests that this does not happen
uniformly.
The light produced is collected by telescopes, recorded by
instruments,

photographic plates or CCD's.
Finally, the astronomer applies further selection criteria to the
images

she obtains in order to create a catalog of galaxies or of clusters
of

galaxies.
By \eq(\ref{eq:composition}), the end result is some effective

transformation, likely to be different for each different category

of objects observed.

Previous models that attempt to relate the statistical properties of
biased

galaxy and matter distributions (Kaiser 1984, Politzer \& Wise 1985,

Bardeen \etal 1986, Szalay 1988, and references therein) have assumed

Gaussian underlying matter fluctuations.
The basic assumption in all these models is that the physical
processes

involved in galaxy formation can be described by a transformation of
the

matter field, $\delta_r(\x)$ smoothed over a galactic scale $r$.
In the notation by Szalay (1988), a local transformation of

the matter fluctuations, $\delta$, leads to the galaxy fluctuations,

$ \delta_g = f(\delta) = G(y)-1 $, where $ y = \delta/\sigma $ is a

normalized matter fluctuation and $G$ is the `luminosity density.'
Equation (\ref{eq:S_g}) shows that assuming the matter fluctuations
are

Gaussian is inadequate for a gravitationally evolved field: at each
order,

the terms arising from gravity and from bias are of comparable
amplitude.
Fry (1986) has considered the case of biasing from
hierarchical matter fluctuations, but only up to the three-point

galaxy correlation function, which was also found hierarchical.

In this paper we have considered the more general case of
hierarchical,

rather than Gaussian, matter fluctuations.
We have shown that any sequence of local biasing transformations

gives a contribution comparable to that from nonlinear gravitational

evolution at each order in $ \xibar_2 $.
As argued in $\S$2.3, this outcome, that a very general nonlinear
bias

preserves the hierarchical structure in the limit of small $ \xi_2 $,

involves a remarkable cancelation which results from the statistical

properties of connected moments.
In a sense gravity evolution for large scales is similar to local
biasing,

but as pointed out by Fry (1984$b$) the self similar time evolution
is a

unique feature of gravity, and what might serve to distinguish
gravity

from any other transformation is the characteristic values of the
amplitudes

$S_j$, which can be calculated explicitly in gravitational
instability.
This, in turn, can allow us to determine properties of the bias
function

from $ S_{g,j} $.

To what extent can we say that the observed galaxy properties

are a consequence of the initial Gaussian conditions?
If we consider gravitational evolution in perturbation theory
the problem of the initial conditions is very simple.
If the initial correlations are in leading order

$ \lexp \delta^j \rexpc  \propto  \lexp \delta^2 \rexp ^\alpha $,

\begin{itemize}

\item $ \alpha < j/2 $  implies non-Gaussian and non-hierarchical

initial conditions that dominate the evolution during the regime
in which $ \lexp \delta^2 \rexp $ is small.

\item $ j/2 < \alpha < j-1 $ implies quasi-Gaussian but
non-hierarchical

initial conditions.
In this case, evolution will produce two contributions to $ \xi_j \,
$:

a dominant non-hierarchical term that grows as $ A^j(t) $ and a

hierarchical term  with characteristic amplitude $S_j$ that grows as

$ A^{2(j-1)}(t) $ but may not become significant until $ \xi \sim 1
$.

\item $ \alpha > j-1  $ implies strongly-Gaussian initial conditions.
In this case, the leading order effect of evolution will produce

hierarchical statistics with characteristic amplitudes $S_j$

for all times.

\end{itemize}
\noindent

That is, the initial conditions for large-scale structure formation

could be Gaussian if, and only if, the evolved matter distribution is

observed to be hierarchical at large scales.

The explicit relations between galaxy and matter amplitudes presented
in equation (\ref{eq:S_g}) show that if we allow biasing to be an
arbitrary function, then the observed galaxy amplitudes can be
arbitrarily

different from the matter ones.
On the other hand, one can use these relations to learn about biasing
by

comparing galaxy amplitudes with theoretical matter predictions.
It is also clear from (\ref{eq:S_g}) that a linear biasing
approximation

is consistent only for the two-point correlation function.
In general, even in the limit of weak fluctuations on very large
scales,

the $j$-point galaxy amplitudes have biasing contributions not only
from
the linear term but from all orders up to $j-1$.

Are the observed hierarchical properties of the galaxy distribution

a consequence of the hierarchical properties of matter?

Or, are they an accident or conspiracy of galaxy-matter biasing?
We have shown here that in the case of local biasing, the observed
galaxy
 hierarchy at large scales can only be a consequence of hierarchical

properties of the smoothed matter distribution and thus suggests that

the initial conditions were indeed hierarchical or Gaussian.

\bigskip
\noindent
{\large\bf Acknowledgements}

\bigskip

This work was supported in part by DOE and by NASA (grant NAGW-2381)
at Fermilab.

\newpage
\bigskip

\noindent
{\large\bf References}
\bigskip

\def\pp{\par\parshape 2 0truecm 16.5truecm 1truecm
15.5truecm\noindent}
\def\paper#1;#2;#3;#4; {\pp#1, {#2}, {#3}, #4}
\def\book#1;#2;#3;#4; {\pp#1, {\sl #2} (#3: #4)}
\def\preprint#1;#2; {\pp#1, #2}

\paper Bardeen, J. M., Bond, J. R., Kaiser, N., \& Szalay, A. S.
1986;%
ApJ;304;15;
\paper Bernardeau, F. 1992;ApJ;392;1;
\pp Bouchet, F. R., Davis, M., \& Strauss M. 1992,

in {\sl Proceedings DAEC Workshop}, ed. G. Mamon, in press
\paper Calzetti, D., Giavalisco, M., \& Meiksin, A. 1992;ApJ;398;429;
\paper Davis, M., \& Peebles, P. J. E. 1977;ApJS;35;425;
\paper Dekel, A., \& Aarseth, S.~J. 1984;ApJ;283;1;
\paper Fry, J. N., \& Peebles, P. J. E. 1978;ApJ;221;19;
\paper Fry, J. N. 1984$a$;ApJ;277;L5;
\paper Fry, J. N. 1984$b$;ApJ;279;499;
\paper Fry, J. N. 1985;ApJ;289;10;
\paper Fry, J. N. 1986;ApJ;308;L71;
\paper Gazta\~naga, E. 1992;ApJ;398;L17;
\paper Gazta\~naga, E. \& Yokohama, J. 1993;ApJ;403;450;
\paper Groth, E. J., \& Peebles, P. J. E. 1977;ApJ;217;385;
\paper Goroff, M. H., Grinstein, B., Rey, S. J., \& Wise, M. B. %
1986;ApJ;311;6;
\paper Guzzo, G., Iovino, A., Chincarini, G., Giovanelli, R., \& %
Haynes, M.~P. 1991;ApJ;382;L5;
\paper Hamilton, A. J. S. 1988;ApJ;332;67;
\paper James, G. S. 1955; Biometrika;42;529;
\paper James, G. S., \& Mayne, A.J. 1962; Sankhy\~a;A24;47;
\paper Kaiser, N. 1984;ApJ;284;L9;
\book Kendall, M. G., Stuart, A., \& Ord, J. K. 1987; The Advanced
Theory of Statistics, Vol.1, 5th ed.;New York;Oxford Univ Press;
\paper Meiksin, A., Szapudi, I., \& Szalay, A. S. 1992;ApJ;394;87;
\book Peebles, P. J. E. 1980; The Large-Scale Structure of the
Universe;%
Princeton;Princeton Univ Press;
\paper Politzer, H. D., \& Wise, M. B. 1985;ApJ;285;L1;
\preprint Saunders, W., Rowan-Robinson M., \& Lawrence A.
 1992; to be published in MNRAS;
\paper Strauss, M. A., Davis, M., Yahil, A., \& Huchra, J. P
1992;ApJ;385;421;
\paper Szalay, A. S. 1988;ApJ;333;21;
\paper Szapudi, I., Szalay, A. S., \& Boschan, P. 1992;ApJ;390;350;

\end{document}